\documentclass[a4paper]{article}
\usepackage{amsmath,graphicx,amssymb,booktabs,bm,pifont,multirow}
\usepackage{bbding}
\usepackage{wasysym}
\usepackage[hidelinks]{hyperref}
\newcommand*{\email}[1]{\href{mailto:#1}{\nolinkurl{#1}} }

\usepackage{INTERSPEECH2022}

\title{A Multi-level Acoustic Feature Extraction Framework for Transformer Based End-to-End Speech Recognition}
\name{Jin Li$^{1,2}$, Rongfeng Su$^{1,2}$, Xurong Xie$^{1,2,3}$, Nan Yan$^{1,2}$, Lan Wang$^{1,2}$}
\address{
{\normalsize $^1$ CAS Key Laboratory of Human-Machine Intelligence-Synergy Systems}, \\
{\normalsize $^2$ Guangdong-Hong Kong-Macao Joint Laboratory of Human-Machine Intelligence-Synergy Systems}, \\
{\normalsize Shenzhen Institute of Advanced Technology, Chinese Academy of Sciences, Shenzhen, China,}\\
{\normalsize $^3$ Institute of Software, Chinese Academy of Sciences, Beijing, China}}
\email{\{li.jin, rf.su, nan.yan, lan.wang\}@siat.ac.cn, xurong@iscas.ac.cn}

\begin{document}

\maketitle
\begin{abstract}
Transformer based end-to-end modelling approaches with multiple stream inputs have been achieved great success in various automatic speech recognition (ASR) tasks. An important issue associated with such approaches is that the intermediate features derived from each stream might have similar representations and thus it is lacking of feature diversity, such as the descriptions related to speaker characteristics. To address this issue, this paper proposed a novel multi-level acoustic feature extraction framework that can be easily combined with Transformer based ASR models. The framework consists of two input streams: a shallow stream with high-resolution spectrograms and a deep stream with low-resolution spectrograms. The shallow stream is used to acquire traditional shallow features that is beneficial for the classification of phones or words while the deep stream is used to obtain utterance-level speaker-invariant deep features for improving the feature diversity. A feature correlation based fusion strategy is used to aggregate both features across the frequency and time domains and then fed into the Transformer encoder-decoder module. By using the proposed multi-level acoustic feature extraction framework, state-of-the-art word error rate of 21.7\% and 2.5\% were obtained on the HKUST Mandarin telephone and Librispeech speech recognition tasks respectively.
\end{abstract}

\section{Introduction}
\label{sec:intro}

In recent years, end-to-end based modelling approaches have gained popularity in automatic speech recognition (ASR) community~\cite{amodei2016deep, chen2021developing, chiu2018state, dong2020cif, meng2021internal, miao2020transformer}. End-to-end ASR models simplify traditional hybrid ASR models by using one single deep neural network instead of acoustic, pronunciation and language components, and thus text can be transcribed directly from speech. Previous researches have shown that significant system performance can be obtained from end-to-end ASR models over hybrid ASR models~\cite{chorowski2014end, luscher2019rwth}.

Transformer~\cite{vaswani2017attention} with self-attention mechanism is one of the effective end-to-end ASR architectures. The self-attention mechanism learns the temporal contextual information of the input sequence by applying attention matrices on the input frames. The motivation behind the self-attention mechanism in Transformer is to aggregate long-term dependencies among acoustic features in the encoder and compute the output sequences in parallel. Current Transformer based ASR modelling approaches can be classified into two categories. The first category is single-stream based methods~\cite{chen2021developing, dong2020cif, miao2020transformer, dong2018speech, luo2021simplified, nakatani2019improving}. Most studies of this category focus on exploring new neural network architecture for faster training speed and online decoding, but they ignore the different characteristic representations of the input signal. In contrast, the multi-stream based modelling approaches mainly focus on the front-end feature processing by using appropriate network structures. For example, in order to enrich the feature diversity, Wang et al.~\cite{wang2019stream} designed a hierarchical attention mechanism to dynamically combine the knowledge from parallel
streams. More recently, Chang et al.~\cite{chang2021end} used three different attention layers to acquire spectral and spatial feature representations from multiple source channels. However, since the previous works of multi-stream based approaches use the same resolution spectrograms for each stream as inputs, the intermediate features derived from each stream might have similar representations and lack diverse feature expressions, such as the vocal characteristics for determining different speakers.

To address this issue, this paper proposes a novel multi-level acoustic feature extraction framework for Transformer based ASR models. Inspired by~\cite{toledano2018multi}, this framework consists of two input streams: a shallow stream with high-resolution spectrograms and a deep stream with low-resolution spectrograms. The shallow branch with less network layers is suggested to preserve the detailed information in the audio inputs that is beneficial for the classification of phones or words. The deep branch with more network layers is suggested to capture the utterance-level speaker characteristic information to improve the feature diversity. Inspired by the non-local operations from~\cite{wang2018non}, a feature correlation based fusion strategy is proposed to integrate both features across the frequency and time domains. The fused features are then fed into Transformer encoder-decoder module to improve the ASR system performance.

The main contributions are summarized as below.
\begin{itemize}
\item[1.]{A novel multi-level acoustic feature extraction framework is proposed in this paper. Speaker-related information can be derived from the deep stream to improve the diversity of the feature representations of speech signal.}
\item[2.]{An additional feature correlation based fusion strategy is proposed to integrate two different types of feature expressions across the frequency and time domains.}
\item[3.]{The Transformer based ASR system using the proposed multi-level acoustic feature extraction framework gave state-of-the-art error rates of 21.7\% and 2.5\% on the HKUST Mandarin telephone speech \cite{liu2006hkust} and Librispeech \cite{panayotov2015librispeech} recognition tasks respectively.}
\end{itemize}

The rest of this paper is organized as follows. The methodology about the multi-level acoustic feature extraction framework is proposed in Section~\ref{sec:Methodology}. Experiments and results are presented in Section~\ref{sec:Experiments}. Finally, conclusions and future works are drawn in Section~\ref{sec:Conclusions}.

\section{Methodology}
\label{sec:Methodology}

\subsection{Transformer based ASR model}
\begin{figure}[ht]
	\includegraphics[width=7cm]{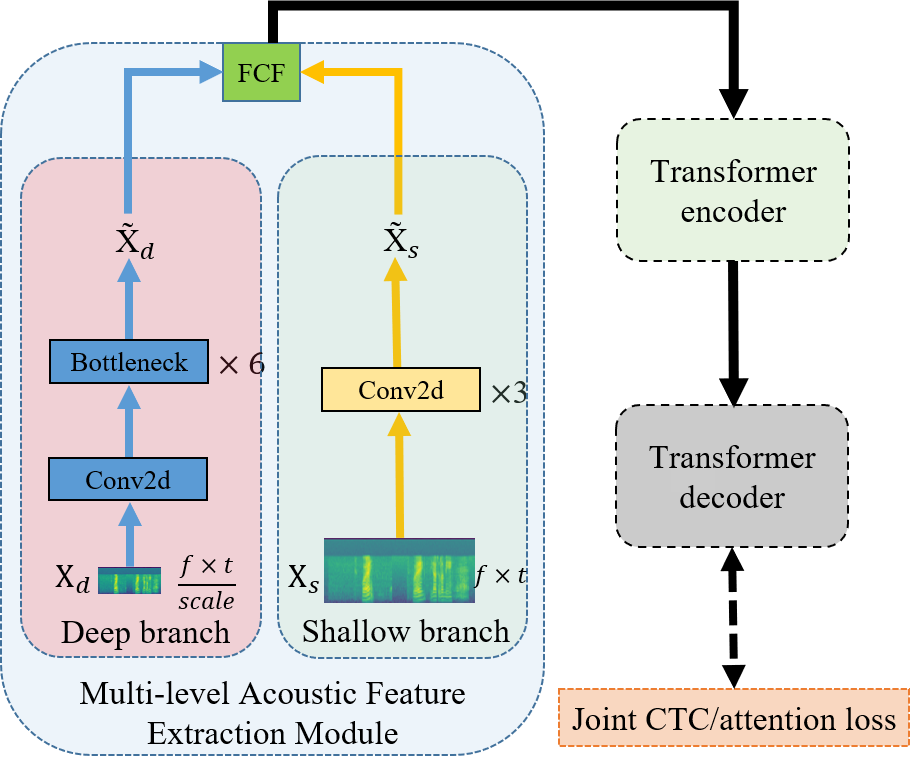}
	\centering
	\caption{Overview of the Transformer based ASR model: ``FCF'' represents the fusion strategy presented in Figure~\ref{fig:fusion}.}
	\label{fig:whole_framework}
\end{figure}

As illustrated in Figure \ref{fig:whole_framework}, the Transformer based ASR model used in this paper consists of the proposed multi-level acoustic feature extraction module, as well as the Transformer encoder and decoder. The multi-level acoustic feature extraction module has two input streams, one is a deep stream with low-resolution spectrograms and the other is a shallow stream with high-resolution spectrograms. After getting features extracted from two streams, they will be integrated by using the fusion strategy proposed in Section~\ref{subsec:Fusion} and then used the inputs of Transformer. Standard Transformer~\cite{vaswani2017attention} containing encoder and decoder components is used in this paper. The encoder is composed of stacked layers that have identical components.
Each layer in the encoder has two sub-layers. The first one is a multi-head self-attention sub-layer and the other is a simple, position wise fully connected feed-forward network with ReLU activation function.
The positional encodings are suggested to use to capture longer dependencies in a sentence.
The decoder structure is similar to the encoder. The only difference is: the decoder inserts a third sub-layer that performs multi-head
attention over the output of the encoder stack. In addition, to enhance the information representations, residual connections around each of the sub-layers are used, which are followed by layer normalization.

\subsection{Multi-level acoustic feature extraction}

There are two streams with different spectrogram resolutions in the multi-level acoustic feature extraction module. Assumed that the input of the shallow stream is $ {\rm X}_s:=(x_1,...,x_T)= \{ x_t \in R^D | t=1,2,...,T \} $ while the input of the deep stream is ${\rm X}_d:=(x_1',..., x_{T'}')= \{ x_t' \in R^{D'} | t=1,2,...,T' \}$, where $T'$ and $T$ ($T'<T$) represent the different scales along the time axis, $D$ and $D'$ are the low and high resolution size of spectrograms respectively. The high resolution spectrogram is suggested to provide the detailed acoustic information especially for the alignment between speech features and labels, which is useful for classifying different phones or words. The low resolution spectrogram is suggested to contain the most important information about the speaker characteristics in each utterance.

\begin{table}[]
	\caption{Neural network configuration in the multi-level acoustic feature extraction module.}
	\label{tab:stream-CNN}
	\centering
	\begin{tabular}{|c|c|c|c|c|}
		\hline
		\multicolumn{1}{|l|}{Stream} & Block  & \multicolumn{1}{l|}{Output Size} & \multicolumn{1}{l|}{Repeats} & \multicolumn{1}{l|}{Stride} \\ \hline
		\multirow{7}{*}{Deep}        & Conv2d    & 32                                   & 1                            & 2                           \\ \cline{2-5}
		& Bottleneck & 32                                   & 1                            & 1                           \\ \cline{2-5}
		& Bottleneck & 32                                   & 1                            & 1                           \\ \cline{2-5}
		& Bottleneck & 48                                   & 3                            & 2                           \\ \cline{2-5}
		& Bottleneck & 64                                   & 3                            & 2                           \\ \cline{2-5}
		& Bottleneck & 128                                  & 2                            & 1                           \\ \cline{2-5}
		& Bottleneck & 256                                  & 2                            & 1                           \\ \hline
		\multirow{3}{*}{Shallow}     & Conv2d    & 128                                  & 1                            & 2                           \\ \cline{2-5}
		& Conv2d    & 256                                  & 1                            & 2                           \\ \cline{2-5}
		& Conv2d    & 256                                  & 1                            & 1                           \\ \hline
	\end{tabular}
\end{table}

The configuration of the neural network for processing the deep and shallow streams is shown in Table~\ref{tab:stream-CNN}.
The neural network for the deep stream contains one 2D convolution block followed by six bottleneck residual blocks from MobileNetV2 \cite{sandler2018mobilenetv2}. The neural network for the shallow stream consists of three 2D convolution blocks. Each convolution block involves batch normalization layer and ReLU activation function.

\subsection{Fusion strategy}
\label{subsec:Fusion}

\begin{figure}[!h]
	\includegraphics[width=5cm]{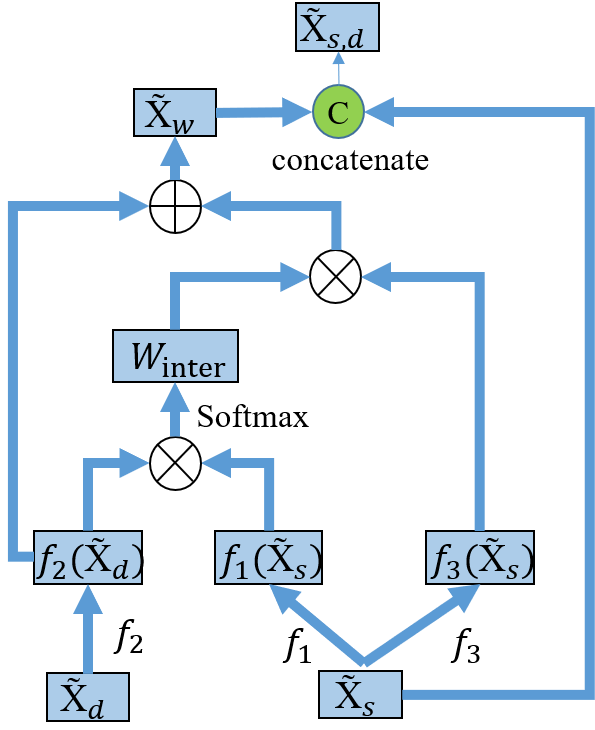}
	\centering
	\caption{Feature Correlation based Fusion (FCF) strategy: The interaction weight $W_\text{\rm inter}$ is used to determine the correlation between the shallow feature $\tilde{\rm X}_s$ and deep feature $\tilde{\rm X}_d$. ``$\otimes$'' represents the dot product operation.}
	\label{fig:fusion}
\end{figure}

As mentioned before, the intermediate features derived from the shallow branch contain the detailed information that is helpful for distinguishing different phones or words, while the outputs of the deep branch contain the information for identifying different speakers.
In order to remove the redundant parts between the shallow and deep features while preserve the complementary ones, a novel feature correlation based fusion strategy shown in Figure \ref{fig:fusion} is used to integrate the both intermediate features across the frequency and time
domains.

Assumed $\tilde{\rm X}_s \in \mathbb{R}^ {c, d_m}$ represents the shallow stream feature vector, $\tilde{\rm X}_d \in \mathbb{R}^ {c, d_m}$ represents the deep stream feature vector rescaled by bilinear interpolation of $\tilde{\rm X}_s$, where $c$ is the number of channels, $d_m = t \times f$ and $f$ is frequency. The interaction weight between the both features is computed as follow.
\begin{equation}
\label{eq:Winter}
	W_\text{inter} := \text{Softmax}(\frac{f_1({\tilde{\rm X}_s})f_2(\tilde{\rm X}_d)^T}{\sqrt{d_m}})
\end{equation}
where $f_1(\cdot)$ and $f_2(\cdot)$ are two different linear transformations with 1D convolution operations, both the outputs of $f_1(\cdot)$ and $f_2(\cdot)$ are two-dimensional matrices, the ``Softmax'' activation function across rows is used. In our early experiments, a weight matrix for $f_3(\cdot)$ could provide faster convergence speed and better system performance. The weighted deep stream feature is computed as follows:
\begin{equation}
	\tilde{\rm X}_w = W_\text{inter} f_3(\tilde{\rm X}_s) + f_2(\tilde{\rm X}_d)
\end{equation}
where $f_3(\cdot)$ has the same definition of $f_1(\cdot)$. The final combined feature is defined as:
\begin{equation}
	\tilde{\rm X}_{s,d} = \text{concat} (\tilde{\rm X}_w, \tilde{\rm X}_s)
\end{equation}
The proposed feature fusion strategy shown in Figure~\ref{fig:fusion} can transmit the speaker-related information into the shallow features and thus enhance the feature diversity.

\subsection{Multi-head self-attention}
The core of Transformer \cite{vaswani2017attention} is utilizing multi-head self-attention mechanism to learn the long-time sequence information.
Multi-head self-attention is designed to integrate the information from different representation subspaces at different positions~\cite{vaswani2017attention}. Each attention head uses the scaled dot-product attention to map a query and a set of key-value pairs to an output. Assumed $Q\in \mathbb{R}^{t_q\times d_q}$ denotes the query matrix, $K \in \mathbb{R}^{t_k\times d_k}$ denotes the key matrix and $V \in \mathbb{R}^{t_v\times d_v}$ denotes the value projection matrix, multi-head self-attention can be computed as:
\begin{equation}
	\mathrm{MultiHead}(Q,K,V)=\mathrm{Concat}({head_{1},...,{head_{h}}})W^O
\end{equation}
\begin{equation}
	head_i=\mathrm{SelfAttn}(QW_i^Q,KW_i^K,VW_i^V)
\end{equation}
\begin{equation}
	\mathrm{Attention}(Q,K,V)=\mathrm{Softmax}(\frac{{QK}^\mathrm{T}}{\sqrt{d_k}} )V
\end{equation}
where the scalar $1/\sqrt{d_k}$ is the normalization term, $head_i$ is the $i^{th}$ head of multi-head self-attention, $W^O$, $W_i^Q$, $W_i^K$ and $W_i^V$ are learnable parameters.

\subsection{Joint CTC/Attention Loss}

The joint CTC/Attention architecture \cite{kim2017joint} profits from both CTC and attention-based models. The objective function is a logarithmic linear combination of the CTC and attention-based objective functions,
\begin{equation}
	\mathcal{L}=\lambda \log P_\mathrm{ctc}(C|\tilde{\rm X}_{s,d}) + (1-\lambda)P_\mathrm{att}(C|\tilde{\rm X}_{s,d})
\end{equation}
where $P_\mathrm{ctc}$ represents CTC loss \cite{graves2006connectionist}, $P_\mathrm{att}$ represents attention loss, $\lambda$ is a trade-off hyperparmeter among CTC and attention loss, which satisfying $0\leq \lambda \leq 1$.

\section{Experiments}
\label{sec:Experiments}
\subsection{Experimental setups}

Experiments were conducted on HKUST Mandarin telephone \cite{liu2006hkust} and Librispeech \cite{panayotov2015librispeech} speech recognition tasks. The HKUST mandarin telephone speech corpus contains about 170 hours training data while 5 hours for the development set and 5 hours for the evaluation set.
The size of the training set of Librispeech corpus is 1000 hours and we used two standard test sets~\cite{panayotov2015librispeech} (``test-clean'' and ``test-other'') for the evaluation.

For both the high-resolution and low-resolution spectrograms, we used 129-dimensional features, including  120-dimensional filterbanks (static, $\Delta$ and $\Delta \Delta$) and 9-dimensional Kaldi pitch (static, $\Delta$ and $\Delta \Delta$). The features were computed within a 25 ms window and the window shift was 10 ms.
The size of the high-resolution spectrogram input ${\rm X}_d$ was 4 times larger than the size of the low-resolution spectrogram input ${\rm X}_s$ by rescaling operation. In addition, we followed the ESPNet \cite{watanabe2018espnet} recipe to preprocess both corpora.

Our experiments were based on the existing recipes in the end-to-end speech processing toolkit ESPNet \cite{watanabe2018espnet}. During the network training, $d_m=256$ in equation (1) was used inside Transformer ASR model. For the joint CTC/Attention loss function, $\lambda = 0.3$ was used. We trained 20 epochs for the HKUST mandarin telephone speech recognition task and 120 epochs for the Librispeech speech recognition task.

\subsection{Experimental results}

\begin{table}[h!]
	\caption{Character error rate (CER) of the Transformer ASR systems only using the shallow or deep stream features, or using both features on the HKUST test set.}
	\label{tab:wostream}
	\centering
	\begin{tabular}{ccc}
		\hline
		Shallow Stream & Deep Stream & CER  \\ \hline
		\checkmark	&             & 23.5 \\
		&    \checkmark         & 58.7 \\
		
		\checkmark	&  \checkmark           & \bf 21.7 \\ \hline
	\end{tabular}
\end{table}

In order to explore the impact of the shallow stream features or deep stream features on the final system performance, we conducted experiments on the Transformer ASR systems with the multi-level acoustic feature extraction module shown in Figure~\ref{fig:whole_framework}.
Table \ref{tab:wostream} shows the performance of the Transformer ASR systems only using the shallow or deep streams, or using both features on the HKUST test set. When only using the deep stream, the character error rate (CER) was up to 58.7\% and it was too weak to use as the output receptive field of the deep stream. This result might be caused by losing lots of acoustic detailed information for distinguishing different phones or words. Moreover, when using the deep stream features as a complementary part for the shallow stream features, the resulted Transformer ASR system outperformed the baseline only using shallow stream features by a CER reduction of 8.3\% relative on the HKUST test set.

\begin{figure}[!h]
	\includegraphics[height=3.5cm]{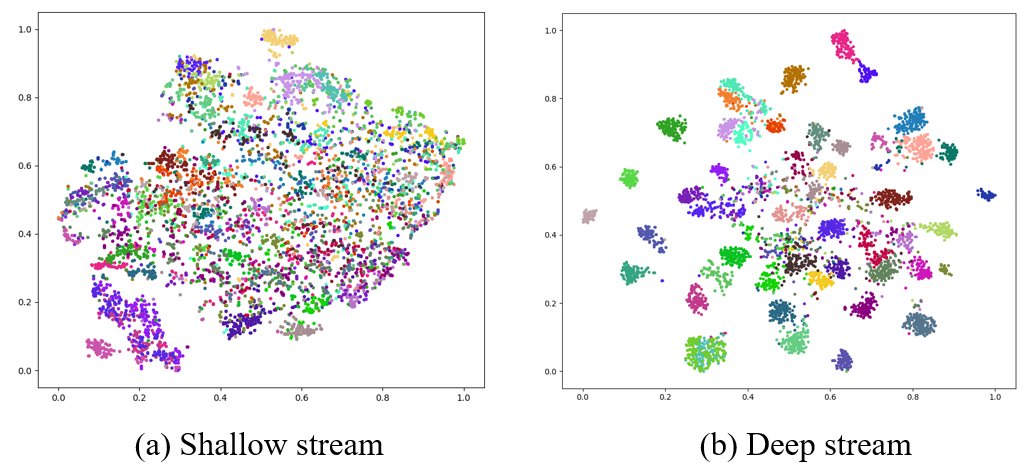}
	\centering
	\caption{The t-SNE~\cite{van2014accelerating} visualization for the (a) shallow stream features and (b) deep stream features on the HKUST test set: different colors represent different speakers. T-SNE with default settings in the sklearn package~\cite{pedregosa2011scikit} is applied for both streams.}
	\label{fig:tsne}
\end{figure}
To further explore the hidden information of the shallow and deep branches in the multi-level acoustic feature extraction module, we used t-SNE~\cite{van2014accelerating} projection to visualize the distribution of the utterance-level shallow and deep stream features. As shown in Figure~\ref{fig:tsne}, since the distance between two arbitrary feature points among the same category is small while the center point of two arbitrary different categories have a far way feature distance, the deep stream features could be used to identify different speakers. It indicates that such features contain the inherent characteristics of vocal cords of different speakers, which are missing in the shallow stream features and might be helpful for speech recognition tasks.

\begin{table}[h!]
	\caption{CER of the Transformer based ASR systems using different fusion strategies on the HKUST test set.}
	\label{tab:fusion}
	\centering
	\begin{tabular}{lc}
		\hline
		Fusion strategy & CER  \\ \hline
		Concatenation & 22.2 \\
		Addition      & 21.9 \\ \hline
		\bf FCF(Ours)     & \bf 21.7 \\ \hline
	\end{tabular}
\end{table}
Table \ref{tab:fusion} shows the CER performance of the Transformer ASR systems using different fusion strategies on the HKUST test set. As shown in Table \ref{tab:fusion}, compared with the ASR systems using the ``concatenation'' and ``fusion'' strategies, the Transformer ASR system using the proposed FCF fusion strategy had the best performance. This indicated that the FCF fusion strategy across the frequency and time axes might be more suitable for capturing the hidden complementary information between the shallow and deep stream features.

\begin{table}[!h]
	\caption{The effect of the number of the bottleneck residual blocks used on the deep branch in the multi-level acoustic feature extraction module on the HKUST Mandarin telephone speech recognition task.}
	\label{tab:num-blocks}
	\centering
	\begin{tabular}{ccc}
		\hline
		\#blocks & development set & test set  \\ \hline
		4      & 22.4       & 22.1 \\
		5      & 22.3       & 21.9 \\
		\bf	6      & \bf 22.2       & \bf 21.7 \\ \hline
	\end{tabular}
\end{table}
Table \ref{tab:num-blocks} shows the Transformer ASR system performance with respect to the number of the bottleneck residual blocks used on the deep branch on the HKUST Mandarin telephone speech recognition task. The number of the bottleneck residual blocks ranged from 4 to 6. As the number of the bottleneck residual blocks increases, the corresponding CER decreases on both the HKUST development and test sets. It would be suggested that more robust speaker-related information can be obtained at a deeper layer.

\begin{table}[!h]
	\caption{Comparison with different state-of-the-art ASR systems on the
		HKUST Mandarin telephone speech recognition task.}
	\label{tab:hkust-result-tune}
	\centering
	\begin{tabular}{l l l}
		\toprule
		\multicolumn{1}{c}{\textbf{Model}} &
		\multicolumn{1}{c}{\textbf{CER}} \\
		\midrule
		Chain-TDNN \cite{povey2016purely}  & 23.7~~~             \\
		Self-attention Aligner \cite{dong2019self}  & 24.1~~~               \\
		CIF \cite{dong2020cif}  & 23.1~~~       \\
		D-Att \cite{gao2021non} & 23.3~~~ \\
		\midrule
		\bf Ours & \bf 21.7~~~       \\
		\bottomrule
	\end{tabular}
\end{table}

We also compared the proposed Transformer ASR system with other state-of-the-art ASR systems on the HKUST Mandarin telephone speech recognition task. As shown in Table~\ref{tab:hkust-result-tune}, compared with other ASR systems, CER reductions of 1.4\%-2.4\% absolute (6\%-10\% relative) were obtained from the proposed speech recognition system on the HKUST corpus.


\begin{table}[!h]
	\centering
	\caption{Experiments in word error rate (WER) on the Librispeech speech recognition task. The ESPNet \cite{watanabe2018espnet} was reimplemented with the setting adim=256, which is comparable with the proposed Transformer based ASR system with $d_m=256$.}
	\begin{tabular}{lcc}
		\hline
		\multicolumn{1}{c}{ASR systems} & test-clean & test-other \\ \hline
		RWTH \cite{luscher2019rwth}                        & 3.8        & 8.8        \\
		DEJA-VU \cite{tjandra2020deja}                     & 2.9        & 6.7        \\
		MEL-t-Fusion-Late \cite{lohrenz2021multi}           & 3.3        & 7.2        \\
		ESPNet \cite{watanabe2018espnet}                      & 3.1        & 6.4        \\ \hline
		{\bf Ours}                         & {\bf 2.5}        & {\bf 5.8}        \\ \hline
	\end{tabular}
\label{tab:librispeech}
\end{table}
We also evaluated our framework on the Librispeech speech recognition task. As shown in Table~\ref{tab:librispeech}, the proposed Transformer ASR system using the multi-level acoustic feature extraction module can achieve state-of-the-art performance on the Librispeech speech recognition task. Especially, compared with ESPNet, our final ASR system gave the WER reductions of 19.4\% relative and 9.4\% relative on the ``test-clean'' and ``test-other'' sets respectively.

\section{Conclusions}
\label{sec:Conclusions}

This paper proposed a novel multi-level acoustic feature extraction framework that can be easily applied into Transformer based models. The proposed framework contained a shallow stream with less network layers and a deep stream with more network layers.
Utterance-level speaker-invariant information derived from the deep stream with low-resolution spectrograms was used to improve the feature diversity, and then a novel feature correlation based fusion strategy was used to integrate the shallow and deep stream features across the frequency and time domains. Experimental results showed that the Transformer based ASR system using the fused features gave state-of-the-art performance on both HKUST Mandarin telephone and Librispeech speech recognition tasks. Future works will focus on applying the proposed feature extraction framework into other ASR models.

\section{Acknowledgements}
This work was supported in part by  the National Key R\&D Program of China (2020YFC2004100), National Natural Science Foundation of China (NSFC U1736202, NSFC 61771461, NSFC 62106255), Shenzhen Peacock Team Project (Grant No. KQTD20200820113106007) and Shenzhen Science and Technology Program (Grant No. JCYJ20210324115810030).

\bibliographystyle{IEEEtran}

\bibliography{mybib}


\end{document}